\documentclass[10pt,final,journal,twocolumn]{IEEEtran}

\ifCLASSINFOpdf

\else

\fi

\usepackage{cite}
\usepackage[cmex10]{amsmath}
\usepackage{amsmath}
\usepackage{algorithmic}
\usepackage{stfloats}
\usepackage{multicol,multienum}
\usepackage{multirow}
\usepackage[dvips]{graphicx}

\usepackage[tight,footnotesize]{subfigure}
\usepackage{wrapfig}
\usepackage{color}

 \usepackage{booktabs}
 \usepackage{multirow}

\usepackage{mathtools}
\usepackage{color}
\usepackage{float}
\usepackage{bm}

\usepackage{pifont}

\usepackage[ruled,linesnumbered]{algorithm2e}

\usepackage{bigfoot}

\begin{document}

\title{Exploring Equilibrium Strategies in Network Games with Generative AI}


\author{Yaoqi~Yang,
Hongyang~Du,
Geng~Sun,
Zehui~Xiong,~\IEEEmembership{Senior Member,~IEEE},
Dusit~Niyato,~\IEEEmembership{Fellow,~IEEE},
and Zhu~Han,~\IEEEmembership{Fellow,~IEEE}

\thanks{Manuscript received xxx. }

\thanks{Yaoqi~Yang is with the College of Communications Engineering, Army Engineering University of PLA, Nanjing 210000, China. (e-mail: yaoqi$\_$yang@yeah.net)}

\thanks{Hongyang Du and Dusit Niyato are with the School of Computer Science and Engineering, NTU, Singapore. (e-mail: hongyang001@e.ntu.edu.sg; dniyato@ntu.edu.sg)}

\thanks{Geng Sun is with the College of Computer Science and Technology, Jilin University, Changchun 130012, China, and also with the College of Computing and Data Science, Nanyang Technological University, Singapore 639798 (email: sungeng@jlu.edu.cn).}

\thanks{Zehui Xiong is with the Information Systems Technology and Design Pillar, Singapore University of Technology and Design, Singapore. (e-mail: zehui$\_$xiong@sutd.edu.sg)}

\thanks{Zhu Han is with the Department of Electrical and Computer Engineering at the University of Houston, Houston, TX 77004, USA. (e-mail: hanzhu22@gmail.com)}
}

\maketitle

\begin{abstract}
Game theory offers a powerful framework for analyzing strategic interactions among decision-makers, providing tools to model, analyze, and predict their behavior. However, implementing game theory can be challenging due to difficulties in deriving solutions, understanding interactions, and ensuring optimal performance. Traditional non-AI and discriminative AI approaches have made valuable contributions but struggle with limitations in handling large-scale games and dynamic scenarios. In this context, generative AI emerges as a promising solution \textcolor[rgb]{0.00,0.00,0.00}{because of} its superior data analysis and generation capabilities. This paper comprehensively summarizes the challenges, solutions, and outlooks of combining generative AI with game theory. We \textcolor[rgb]{0.00,0.00,0.00}{start with} reviewing the limitations of traditional non-AI and discriminative AI approaches in employing game theory, and then highlight the necessity and advantages of integrating generative AI. \textcolor[rgb]{0.00,0.00,0.00}{Next}, we explore the applications of generative AI in various stages of the game theory lifecycle, including model formulation, solution derivation, and strategy improvement. Additionally, from game theory viewpoint, we propose a generative AI-enabled framework for optimizing machine learning model performance against false data injection attacks, supported by a case study to demonstrate its effectiveness. Finally, we outline future research directions for generative AI-enabled game theory, paving the way for its further advancements and development.
\end{abstract}

\begin{IEEEkeywords}
Generative AI, game theory, game theoretical model formulation, equilibrium solution derivation.
\end{IEEEkeywords}

\IEEEpeerreviewmaketitle

\section{Introduction}


Game theory is a crucial tool in mathematics and economics, aiding individuals in strategic decision-making. It provides a solid framework for finding the equilibrium solutions by analyzing how actions and reactions interact among participants. While game theory is excellent at predicting rational behavior and considering factors like incentives and information dynamics, it does have \textcolor[rgb]{0.00,0.00,0.00}{the following} limitations, \textcolor[rgb]{0.00,0.00,0.00}{and therefore addressing} these limitations is important for its ongoing development and wider use\footnote{https://www.iienstitu.com/en/blog/game-theory-strategic-analysis-and-practical-applications}.
\begin{itemize}
  \item \emph{The equilibrium solution of the game is difficult to solve}. The solving process requires advanced mathematical techniques and significant computational resources. As games become more complex, either with more players or more sophisticated strategies, the computational requirements increase, \textcolor[rgb]{0.00,0.00,0.00}{thus} making it difficult to derive equilibrium solutions.
  \item \emph{The game process is complicated to understand.} Understanding the game process is complex. The interactions among multiple individuals, each with their own goals and constraints, make analysis difficult. Analyzing why players interact in the way they do requires not just understanding the game's rules but also having accurate insights into the real-world situation.
  \item \emph{The game-based strategy may not lead to the best performance}. Not all players may act rationally, as imperfect information or biased judgements can affect outcomes. Additionally, focusing on individual optimal strategies within the game may not always result in the best overall outcomes for the system or network as a whole, \textcolor[rgb]{0.00,0.00,0.00}{thereby} reducing the effectiveness of the solution.
\end{itemize}

To address these challenges, traditional non-AI methods have been adopted, making significant contributions. For instance, they excel in formulating precise game models, enabling mathematical analysis to uncover the core of strategic decision-making. This approach offers theoretical solutions and assesses the effects of different strategies. Moreover, numerical optimization methods serve as robust computational tools for solving game theory problems, providing solutions when analytical methods are not feasible. However, despite their widespread use, traditional non-AI approaches in game theory still face several \textcolor[rgb]{0.00,0.00,0.00}{following} limitations\footnote{http://article.sapub.org/10.5923.j.jgt.20200902.01.html}.
\begin{itemize}
\item \emph{Limited participants and actions consideration}. When assessing the iterative dynamics between two individuals\footnote{Note that even though mean-field game can be applied in the large scale system, it can describe only relationships between one individual and the mean-field formulated by the rest of individuals. Relationships between any two individuals are still difficult to figure out.} in game-solving processes, traditional non-AI methods typically oversimplify strategic interactions by restricting the number of players and actions considered. Such simplifications hinder the application of game theory in complex systems characterized by numerous interacting individuals and a wide array of potential strategies.
\item \emph{Standard game structure requirement}. Traditional non-AI methods rely on predefined rules and payoffs to analyze games and find equilibrium solutions. However, many strategic environments are dynamic, with changing rules and uncertain payoffs. This can lead to inaccurate predictions or suboptimal strategies due to the rigid structure of traditional games.
\item \emph{Strong participant's behavior assumption}. Traditional non-AI approaches assume strong rationality, where participants make decisions to maximize their own utility and accurately predict others' actions. However, when participants have incomplete or imperfect information, they may act biasedly or non-ideally, invalidating these assumptions.
\end{itemize}

Meanwhile, discriminative AI offers a promising approach to addressing challenges in game theory. Supervised learning models, for instance, can predict participants' behavior based on historical data, improving decision-making by guiding payoff-oriented actions. Reinforcement learning, on the other hand, can help agents learn equilibrium strategies by treating strategic interactions as Markov decision processes, allowing them to converge to solutions over repeated engagements. However, despite its potential, discriminative AI faces challenges as follows\footnote{https://www.emerald.com/insight/content/doi/10.1108/JDAL-10-2021-0011/full/html}.
\begin{itemize}
\item \emph{Demanding a large amount of labeled data}. Training discriminative AI models to identify strategies, predict participants' actions, and select suitable strategies necessitates a vast amount of labeled data to establish the correlation between inputs and outputs. However, this poses challenges in complex and dynamic environments where collecting labeled data is impractical or unfeasible.
\item \emph{Struggling to unseen game types and scenarios}. Discriminative AI models are typically trained on specific datasets, making them struggle to generalize effectively with new game types or scenarios not seen. Retraining these models on new datasets to accommodate unseen scenarios can prove time-consuming and impractical, especially in rapidly changing environments.
\item \emph{Capturing complex dynamics traits of games inefficiently}. Discriminative AI models often rely on fixed representations of input data, which may not adequately capture non-linear or stochastic relationships inherent in game processes. They may also overlook emergent behaviors or strategic patterns, limiting their predictive capability in dynamic scenarios.
\end{itemize}

In this regard, to address limitations in traditional non-AI and discriminative AI approaches, generative AI offers the following effective solutions.

\begin{itemize}
\item \emph{Offering the digital twin evolving environment, modeling the opponent in the game theoretical model formulation stage.} Generative AI can be a powerful tool to realize the digital twin \cite{tao2023wireless}, as it can generate synthetic data to support specific functional models's training and deployment to provide service. Specifically, generative AI can offer a digital twin of the evolving environment, where participants make strategic decisions through interacting with each other. Moreover, generative AI can also help model the opponent when there is limited priori information \cite{chivukula2020game}. In this regard, generative AI can address concerns of data limitation, reduce the dependency of standard game structure, relax the assumption of the participant's behaviors, and be efficient in deriving equilibrium solutions for unseen scenarios.
\item \emph{Creating agents to simulate the game process, using generative models to derive equilibrium solutions in the game problem solution derivation stage.} Generative AI can create intelligent agents \cite{he2024generative} to simulate the participants within the game process. These agents can act as strategic entities within the environment, making decisions based on predefined objectives, learning from past experiences, and adapting to changing circumstances. Through these simulated interactions, participants can gain insights into the dynamics of strategic decision-making, uncovering emergent behaviors, and identifying strategies that lead to desirable outcomes. In addition, generative AI \textcolor[rgb]{0.00,0.00,0.00}{is able to} directly provide equilibrium solutions with the models' outputs \cite{farnia2020gans}. In this regard, generative AI can break through the player and action number limitation, capture the dynamic nature of the game process effectively, obtain equilibrium solutions more efficiently, as well as understand the game process by observing the status and behaviors of the created agents.
\item \emph{Improving deep reinforcement learning-based decision making performances, enhancing the derived equilibrium solution with new game scenarios in the game problem strategy improvement stage.} Generative AI can be used to improve deep reinforcement learning, such an integration can address concerns on modeling complex environments, instability and slow convergence \cite{du2023beyond}. Additionally, by generating diverse environments \cite{ebert2023generative} with varying parameters and constraints, participants can evaluate the robustness and effectiveness of derived strategies. This iterative process leads to the development of more resilient and adaptive strategies across various environments, significantly enhancing game-theoretical strategy performance.
\end{itemize}

\begin{table*}[t]
\centering
\caption{Comparison for traditional non-AI, discriminative AI and generative AI-based game theory solving schemes. } \label{comparison1}
\resizebox{0.85\width}{!}{
\begin{tabular}{c}
\includegraphics[width=18cm]{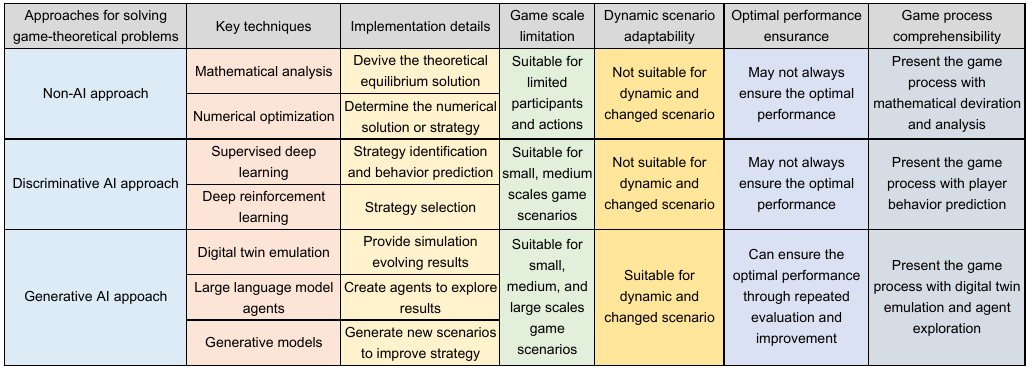}\\
\end{tabular}}
\end{table*}

Based on the analysis above, we summarize and compare traditional non-AI, discriminative AI, and generative AI-based game theory solving approaches in Table I. The integration of generative AI and game theory not only overcomes limitations faced by game theory but also effectively addresses concerns with traditional non-AI and discriminative AI methods. \textcolor[rgb]{0.00,0.00,0.00}{Different from our previous work that focuses on utilizing retrieval-augmented generation (RAG) and large language model (LLM) to solve game theoretical problems \cite{he2024generative}, this paper extends the generative AI's applications for the whole life cycle of game theory, including game model formulation, solution derivation, and strategy improvement.} Specifically, the key contributions are summarized as follows.
\begin{itemize}
\item  We introduce the preliminaries for game theory and generative AI. Through a comprehensive review of these two techniques, we clarify the motivations and advantages of integrating generative AI into the game-theoretical problem-solving process. Additionally, we highlight the aspects and manner in which the generative AI technique can outperform traditional non-AI and discriminative AI-based  approaches.
\item We present how to combine generative AI with game theory, focusing on the whole life cycle of game theory, model formulation, solution derivation, and strategy improvement. In each stage, we detail the services implementation outlines and advantages of generative AI. Furthermore, we illustrate the potential challenges of integrating generative AI into game theory.
\item We propose a generative AI-enabled game theoretical framework for optimizing machine learning (ML) model's performance under false data injection attack, and a mobile crowdsensing-based vehicle image collection is designed as a case study. Experiments results have demonstrated that by generating synthetic data, Generative Adversarial Networks (GAN) and Generated Diffusion Model (GDM) can model opponents and generate new game scenarios, and LLM (i.e., Transformer-based model) can formulate game theoretical problems and create agents to explore game results.
\end{itemize}

The rest of the paper is organized below. Section II introduces the preliminaries of game theory and generative AI. Then, Section III presents the generative AI-enabled game theoretical applications, where potential challenges in integrating generative AI into game theory are also discussed. Besides, under false data injection attack, the framework for optimizing ML model's performance with generative AI-enabled game theoretical approach is detailed in Section IV. Next, Section V highlights several future research directions. Finally, Section VI concludes the paper.

\section{Preliminaries of game theory and generative AI}

\subsection{Preliminaries of game theory }

Game theory offers valuable insights into decision-making when individual choices impact others' outcomes. A complete game process comprises three main components: players, actions, and utility functions.

\emph{Player.} Players are individuals actively engaged in the game. To establish player modeling at the game's onset, various discriminative AI techniques can be employed. For instance, \cite{collins2021interactive} utilizes intelligent agent-based modeling and simulation (ABMS) to depict player behavior. This approach is particularly applied in cooperative game scenarios within strategic coalition formation contexts. Numerical findings indicate that the ABMS method aligns well with player behavior, with 42\% of trials resulting in outcomes predicted by the underlying theory of the proposed scheme. Nonetheless, discriminative AI-driven player modeling may encounter challenges in adapting to game environment changes or opponent behavior, thereby limiting effectiveness in dynamic or intricate game settings.

\emph{Action.} Actions represent the decisions made by players. In assessing player actions during interactions, \cite{kusyk2021artificial} has examined the efficacy of a discriminative AI-based approach. Specifically, a bio-inspired evolutionary method is employed to ascertain the targeted movement positions of a UAV swarm. This approach generates candidate positions that serve as strategies in a realistic, self-enforcing non-cooperative game scenario involving a UAV and its nearby counterparts. However, discriminative AI-based action determination schemes may encounter challenges in generalizing across diverse game scenarios or opponent profiles, thereby diminishing performance in unforeseen circumstances.

\emph{Utility function.} The utility function assigns a value to each potential outcome of the game for every player, reflecting their preferences. To maximize individual player's payoffs through utility function design, \cite{jiao2013novel} proposes a mathematical mixed model that reflects players' benefits and objectives. This approach concurrently considers player benefits and pricing, thereby enhancing Pareto payoff outcomes. Nonetheless, traditional non-AI-based utility function design schemes may struggle to capture characteristics of complex game environments, potentially resulting in oversimplification or insufficient utility representation.

To overcome limitations of traditional non-AI and discriminative AI approaches in player modeling, action determination, and utility function design, we integrate generative AI into game theory. This integration aims to provide new insights into game theory's design, solving, and implementation.

\subsection{Preliminaries of generative AI}

After being trained on extensive datasets of existing samples, generative AI has the capability to generate novel data or content from scratch. This methodology learns to recognize patterns and structures within the samples, thereby producing synthetic data that is similar but not identical. Commonly employed generative AI models encompass Variational autoencoder (VAE), Flow-based model (FBM), Generative adversarial network (GAN), Generative diffusion model (GDM), and Transformer-based model (TBM) \cite{cao2023comprehensive}.

\emph{Variational autoencoder (VAE).} VAE comprises an encoder and a decoder. The encoder maps input data into a latent space, while the decoder reconstructs the data from this latent space. VAE is trained to minimize reconstruction error while also regularizing the latent space to adhere to a specific distribution. This regularization encourages VAE to generate new data points similar to those in the training set, facilitating smooth interpolation in the latent space. VAE finds applications in realizing game-theoretical adversarial deep learning scenarios. For instance, in achieving game-theoretical learning between a variational adversary and a Convolutional Neural Network (CNN), \cite{chivukula2020game} designs a variable-sum two-player sequential Stackelberg game. VAE is then employed to determine the adversary¡¯s payoff function based on data manipulation. Numerical results demonstrate that the proposed method achieves a classification error of 6.44\%, whereas the original performance without defense stands at 10.01\% classification error.

\emph{Flow-based model (FBM).} FBM learns the data distribution by converting a simple base distribution into the target distribution via a sequence of invertible transformations. It accurately computes data likelihoods, rendering it valuable for tasks emphasizing likelihood estimation, like density estimation or anomaly detection. Specifically, FBM can be used in the multi-agent continuous control scenario \cite{luo2024multi}. This application enables FBM to function as a solution tool for game-theoretical problems and to provide guidance for strategic decision-making by players.

\emph{Generative adversarial network (GAN).} GAN consists of a generator and a discriminator. The generator produces synthetic data samples, while the discriminator endeavors to differentiate between real and synthetic samples. Through adversarial training, the generator learns to generate samples that are indistinguishable from real data as perceived by the discriminator. Additionally, GAN can offer a solution to the zero-sum game known as the proximal equilibrium \cite{farnia2020gans}. Unlike the Nash equilibrium, the proximal equilibrium captures the sequential nature of GAN, where the generator acts first, followed by the discriminator. Numerical findings demonstrate that the highest sample quality can be achieved at the proximal equilibrium point.

\emph{Generative diffusion model (GDM).} GDM generates samples by progressively applying noise to an initial data sample. This iterative process begins with a clean data sample and incrementally introduces noise over multiple steps until producing the final sample. Moreover, from a hypergraph game perspective, when the channel allocation problem for ultradense cloud D2D communication networks with asymmetric interference is modeled as a combinational optimization problem\footnote{https://ieeexplore.ieee.org/document/8361901}, GDM can effectively solve it by employing graph-based denoising diffusion models to generate high-quality solutions \cite{sun2024difusco}.

\emph{Transformer-based model (TBM).} TBM utilizes self-attention mechanisms to handle sequences of data. It achieves data generation by conditioning the process on a prompt or a set of features, then generating new data samples based on this conditioning. Additionally, TBM aids in game theory by assisting players in determining actions to maximize payoff. For instance, TBM can be integrated into reinforcement learning\footnote{https://arxiv.org/pdf/2301.03044}, offering decision-making support. Through this approach, improved performance can be attained with derived strategies.

Moreover, Table \ref{comparison2} provides a summary of typical generative AI models, encompassing their components, advantages, drawbacks, offered game-theoretical solutions, and potential applications in wireless systems.

\begin{table*}[t]
\centering
\caption{Summarization of typical generative AI models, provided game-theoretical solutions, and possible applications. } \label{comparison2}
\resizebox{0.9\width}{!}{
\begin{tabular}{c}
\includegraphics[width=18cm]{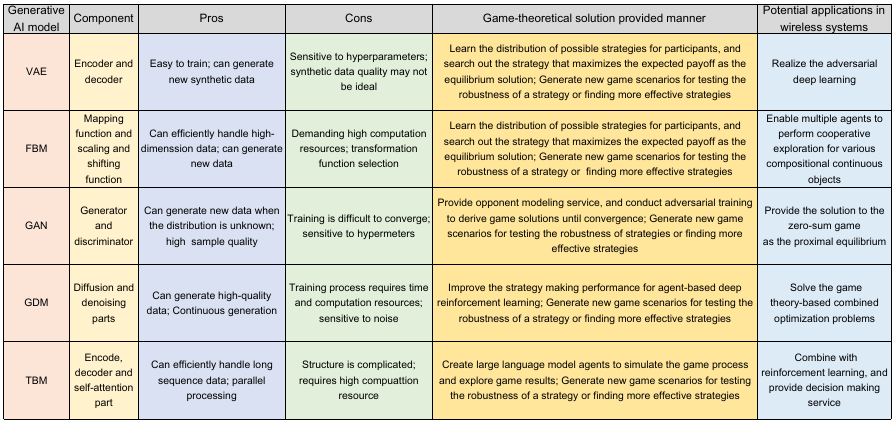}\\
\end{tabular}}
\end{table*}

\section{Generative AI-enabled game-theoretical applications}

\subsection{Potential application scenarios}

From the viewpoint of \emph{model formulation}, \emph{solution derivation}, and \emph{strategy improvement} aspects of game theory's whole life cycle, we explore the issues of generative AI-enabled game-theoretical applications, which are summarized as below.

\subsubsection{Generative AI for game theoretical model formulation}\

By incorporating participant and environmental elements, a game theory model can be established. Generative AI plays a crucial role in this process by contributing to providing emulation environments and opponent modeling. This enables the exploration of complex interactions and strategies among players in a variety of game settings.

\emph{Providing the emulation environment to evaluate the game process}. Generative AI facilitates the creation of a digital twin simulation platform, which simulates various game scenarios based on predefined rules and player actions. This platform enables players to analyze the outcomes of different strategies and decisions within the game. For instance, in the context of 6G wireless networks, generative AI serves as a crucial enabler for realizing digital twins \cite{tao2023wireless}. Specifically, \cite{tao2023wireless} proposed a hierarchical generative AI-enabled framework for wireless network digital twins at both the message-level and policy-level. This framework offers emulation, evaluation, and optimization services for physical entities (e.g., players of one game). Experiments results have demonstrated that when training AI models to realize admission control in the network slicing aspect, the proposed generative AI-enabled model is with 0.999 accuracy, while the Long Short-Term Memory (LSTM) model can only reach 0.992 accuracy.

\emph{Serving as the potential opponent to establish the game model}. Generative AI proves invaluable in modeling opponents' behavior within a game. By discerning the probable actions of adversaries, a player can adjust their strategy to gain a competitive advantage. This dynamic interaction greatly enhances the game's dynamics and boosts player payoffs. For instance, deep learning-based Spear Phishing attack defense schemes often neglect attacker models, potentially undermining defense efficacy. To address such a concern, \cite{kamran2021semi} utilizes GAN to design games between the attacker and defender in training and deployment phases. Numerical results have shown that the proposed scheme can realize 0.9552 Area Under Curve (AUC) on Phishing uniform resource locator (URL) classification performance, while baselines Support Vector Machine (SVM), CNN, and LSTM are with 0.8638, 0.9251, and 0.9471 AUC values, respectively.

\subsubsection{Generative AI for game-theoretical  solution derivation}\

When addressing game theory problems, the quality of the derived solution directly impacts players' payoffs and system performance. Generative AI, in this context, can create agents to explore game outcomes, as well as in deriving equilibrium solutions and strategies. This contribution streamlines the solving process and enhances comprehensibility by observing agents' behaviors.

\emph{Creating agents to simulate game process and explore game results}. Trained generative AI models can be utilized to create agents that realistically emulate players' behaviors within a game environment. Once deployed in simulated game environments, these agents interact with each other and the environment until reaching an equilibrium point. \textcolor[rgb]{0.00,0.00,0.00}{For instance, \cite{he2024generative} proposed a framework leveraging LLMs to create agents capable of reasoning and decision-makings, providing final response to the input requirement. Numerical results with a UAV secure communication optimization case indicate that generative AI algorithm within such a framework can reach the same performance as the expert scheme.}


\emph{Using generative model to derive equilibrium solutions and strategies}. Once the generative AI model is trained, it becomes a valuable tool for analyzing equilibrium solutions and strategies within games. This process entails inputting various initial conditions or player strategies into the model and observing the outcomes that it produces. Through iterative adjustment of input conditions, the model can progressively converge towards equilibrium solutions. For example, to investigate the relationship between generative AI's outcomes and game theory's equilibrium solutions, \cite{zhaogenerative} proposes one probabilistic framework for modeling multi-agent decision-making problems. Within such a framework, \cite{zhaogenerative} establishes the theoretical connection between the flow equilibrium and the Nash equilibrium, aiming to providing equilibrium solutions and strategies with generative AI models. \textcolor[rgb]{0.00,0.00,0.00}{When being adopted to solve the differential game, numerical results have demonstrated the proposed approach can converge to the global optimum, while the baseline Deterministic Policy Gradient (DDPG) can only reach the sub-optimal solution, indicating the superiority of this approach over existing schemes.}


\subsubsection{Generative AI for game-theoretical  strategy improvement}\

Given the dynamic nature of game environments and the challenge of incomplete player observation, game-theoretical strategies and solutions must undergo iterative refinement. In this regard, generative AI can improve deep reinforcement learning, and test solutions in new generated scenarios. This manner can further increase the player's payoffs as well as enhance the system performance.

\emph{Improving deep reinforcement learning-based decision making performances}. When solving game theoretical problems, deep reinforcement learning has become a cornerstone in guiding player strategy and decision-making processes. However, to enhance its performance in modeling complex environments and mitigate issues such as instability and slow convergence, generative AI offers valuable insights. For instance, \cite{du2023beyond} tackles a power allocation problem involving users across multiple orthogonal channels. Employing a generative diffusion model-based reinforcement learning approach, their optimization objective is to maximize the sum rate of all channels.
\textcolor[rgb]{0.00,0.00,0.00}{In this regard, such a scheme can also be implemented in a game theory scenario, where the game-based resource allocation/optimization problems can obtain better performance with diffusion model-based reinforcement learning.}


\emph{Enhancing the derived equilibrium solution in new game scenarios}. Generative AI has the capability to generate an extensive array of new scenarios within complex games, offering players insights into potential outcomes of different strategies and enabling them to refine their decision-making processes. By iteratively testing and refining solutions within these new scenarios, system performance can be further enhanced. For instance, \cite{ebert2023generative} leverages \textcolor[rgb]{0.00,0.00,0.00}{LLMs} to streamline software development processes, automating tasks such as testing, debugging, and deployment. The results demonstrate the potential for generative AI to automate testing procedures, ensuring thorough testing and verification of each requirement. \textcolor[rgb]{0.00,0.00,0.00}{Similarly, the above approach can also be adopted in game theory, where the current equilibrium solution/strategy can be significantly evaluated and further improved under generative AI-based new scenarios.}

\subsection{Lessons learned}

\begin{table*}[t]
\centering
\caption{Services comparison between traditional non-AI/discriminative AI and generative AI techniques within the whole life cycle of game theory.} \label{comparison3}
\resizebox{0.9\width}{!}{
\begin{tabular}{c}
\includegraphics[width=18cm]{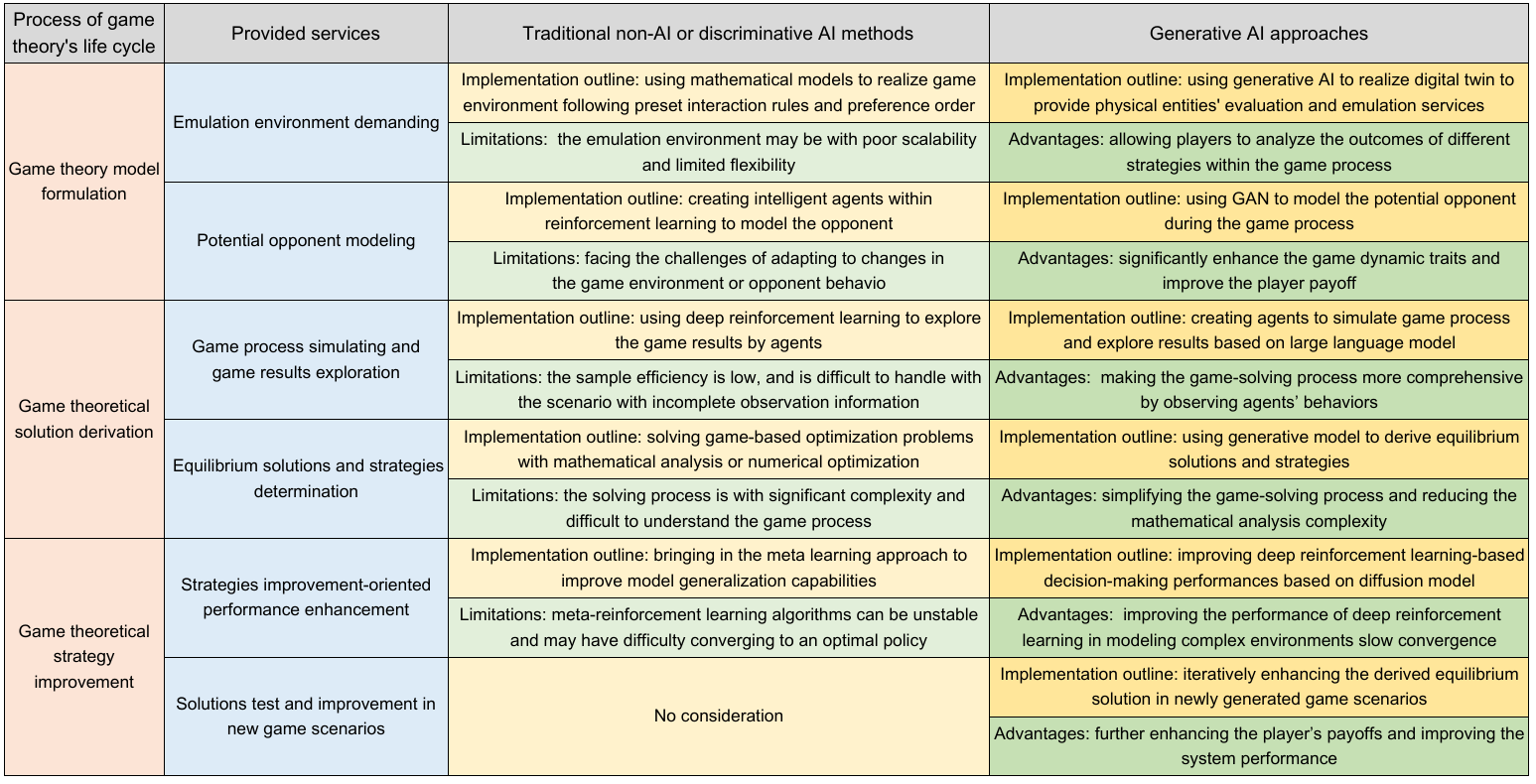}\\
\end{tabular}}
\end{table*}

In summary, generative AI offers substantial insights throughout the entire lifecycle of game theory. A comprehensive comparison of services provided by traditional non-AI/discriminative AI and generative AI techniques is presented in Table \ref{comparison3}. However, despite the efficiency gains and expanded applicability generative AI brings to game theory, its integration faces several challenges.
\begin{enumerate}
\item Training generative AI models demands significant computational resources and energy consumption, potentially limiting its adaptability in resource-constrained scenarios.
\item Generative AI models involve numerous hyperparameters and require substantial storage capacity, posing challenges for implementation in resource-limited environments. 
\item The training and implementation of generative AI in game theory require considerable time, making it difficult to achieve real-time responsiveness in the solving process of generative AI-enabled game theory.
\end{enumerate}

\section{Machine learning model performance optimization under false data injection attack: a generative AI-enabled game theoretical approach}

\subsection{Research motivation}

To train the ML model for object detection (identification, classification), a high-quality dataset is needed. When collecting the required data, honest users contribute genuine data to enhance model performance, while malicious users may launch false data injection attacks by providing synthetic data to degrade model performance \cite{rankin2020reliability}. Since generating synthetic data and collecting real data incur different costs, designing data collection/generation strategies for both honest and malicious users is of paramount importance. Here, there are natural conflict and confrontation relationships among honest and malicious users, game theory can be considered to solve above problems. However, directly applying game theory to address these issues is challenging. For example, malicious users are difficult to model with limited prior knowledge, the game process is complicated to analyze in the dynamic environment, and the equilibrium solution is challenging to derive with optimal performance. In this regard, we propose a generative AI-enabled game theoretical framework. This framework aims to optimize ML model performance under false data injection attack. It facilitates the entire lifecycle of game theory with generative AI, encompassing model formulation, solution derivation, and strategy evaluation.

\subsection{Proposed framework}

\begin{figure*}[!htb]
  \centering
  \includegraphics[width=18cm]{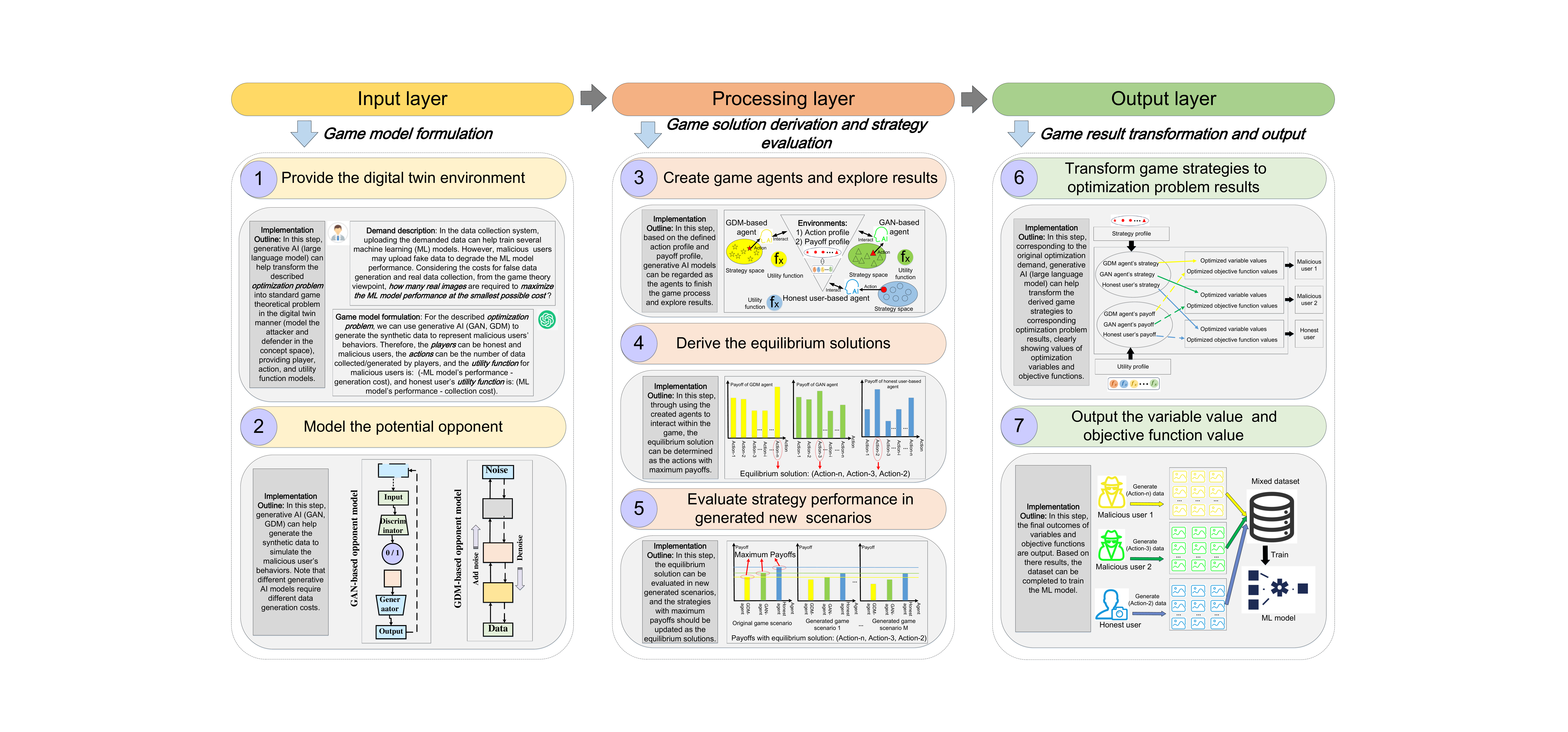}\\
  \caption{The generative AI-enabled game theoretical framework for optimizing ML model's performance under false data injection attack. The framework follows a layered architecture, including input layer, processing layer, and output layer. The input layer can formulate game model, providing the digital twin environment and modeling the potential opponent. The processing layer can derive the game solution and evaluate the strategy, providing services on creating game agents and exploring results, deriving equilibrium solutions, and evaluating strategy performance in generated new scenarios. The output layer can transform and output the game results, mainly focusing on transforming game strategies to optimization problem results, and outputting values of variables and objective functions. }
  \label{GAIMODEL}
\end{figure*}

To efficiently optimize performance of the ML model against false data injection attacks while minimizing costs, we propose a layered generative AI-enabled game theoretical framework, as depicted in Fig. \ref{GAIMODEL}. This framework comprises an input layer, processing layer, and output layer.

\emph{The input layer is for game model formulation, which can be realized on providing the digital twin environment and modeling the potential opponents two aspects}. From a game theory perspective, generative AI can assist in transforming the optimization problem into a game theoretical one, providing models for players, actions, and utility functions. This involves conceptualizing attackers (malicious users) and defenders (honest users) within the game process, thus creating digital twins from physical entities to virtual models. To further characterize potential attacker behaviors, generative AI can deploy various generative models to simulate different types of attackers, generating diverse synthetic data.

\emph{The processing layer focuses on game solution derivation and strategy evaluation, including creating game agents and exploring results, deriving equilibrium solutions, and evaluating strategy performance in generated new scenarios}. Initially, generative AI models can act as agents to navigate the game process and explore outcomes based on defined action and payoff profiles. By employing these agents to interact within the game, equilibrium solutions can be determined. Subsequently, these equilibrium solutions can be evaluated in newly generated scenarios by generative AI, ensuring strategies with maximum payoffs are updated as ultimate equilibrium solutions.

\emph{The output layer can realize the game results transformation and output them, including transforming game strategies to optimization problem results, and outputting the variable value and objective function value}. In line with the original optimization requirements, generative AI aids in transforming derived game strategies into corresponding optimization problem results, elucidating the values of optimization variables and objective functions.

\subsection{Case study and results analysis}

\emph{Scenario description}. When gathering vehicle image data through mobile crowdsensing, the ML model is trained for downstream tasks such as vehicle identification. However, the presence of malicious sensing users uploading synthetic data can degrade the dataset quality, thus impacting the ML model's performance \cite{rankin2020reliability}. To address this, a game can be formulated between honest and malicious users, where the former aim to enhance the ML model's performance while the latter aim to decrease it. Within a fixed number of image datasets, the optimization of the number of uploaded images by both honest and malicious users is crucial to maximize their respective payoffs, which depend on the ML model's performance and the costs associated with image collection/generation.

\emph{Experiment settings}. The real vehicle image dataset is available at Caltech Computational Vision website\footnote{https://data.caltech.edu/records/f6rph-90m20} created by Pietro Perona. The \textcolor[rgb]{0.00,0.00,0.00}{LLM} agent is based on ChatGPT 4\footnote{https://chat.openai.com/}.  GAN\footnote{https://ww2.mathworks.cn/help/deeplearning/ug/train-generative-adversarial-network.html}, GDM\footnote{https://ww2.mathworks.cn/help/deeplearning/ug/generate-images-using-diffusion.html} are implemented on Matlab R2023a, running on a laptop with Intel(R) Core(TM) i7-11700 CPU @ 1.60GHz, NVIDIA GeForce RTX 3060 GPU, 32.0 GB memory and Windows 11. In addition, the adopted ML model is SVM\footnote{https://ww2.mathworks.cn/help/stats/classificationlearner} to identify the existence of vehicles. Moreover, the total image number of the required dataset is 300, and the image collection/generation costs for honest user, GAN-based malicious user, and GDM-based malicious user are set as 0.0005, 0.0007, and 0.0008 per image. \textcolor[rgb]{0.00,0.00,0.00}{In addition, the source code for this case study is available on github\footnote{https://github.com/Yaoqi-Yang97/GAI4Game}.}

\begin{figure*}[!htb]
  \centering
  \includegraphics[width=11cm]{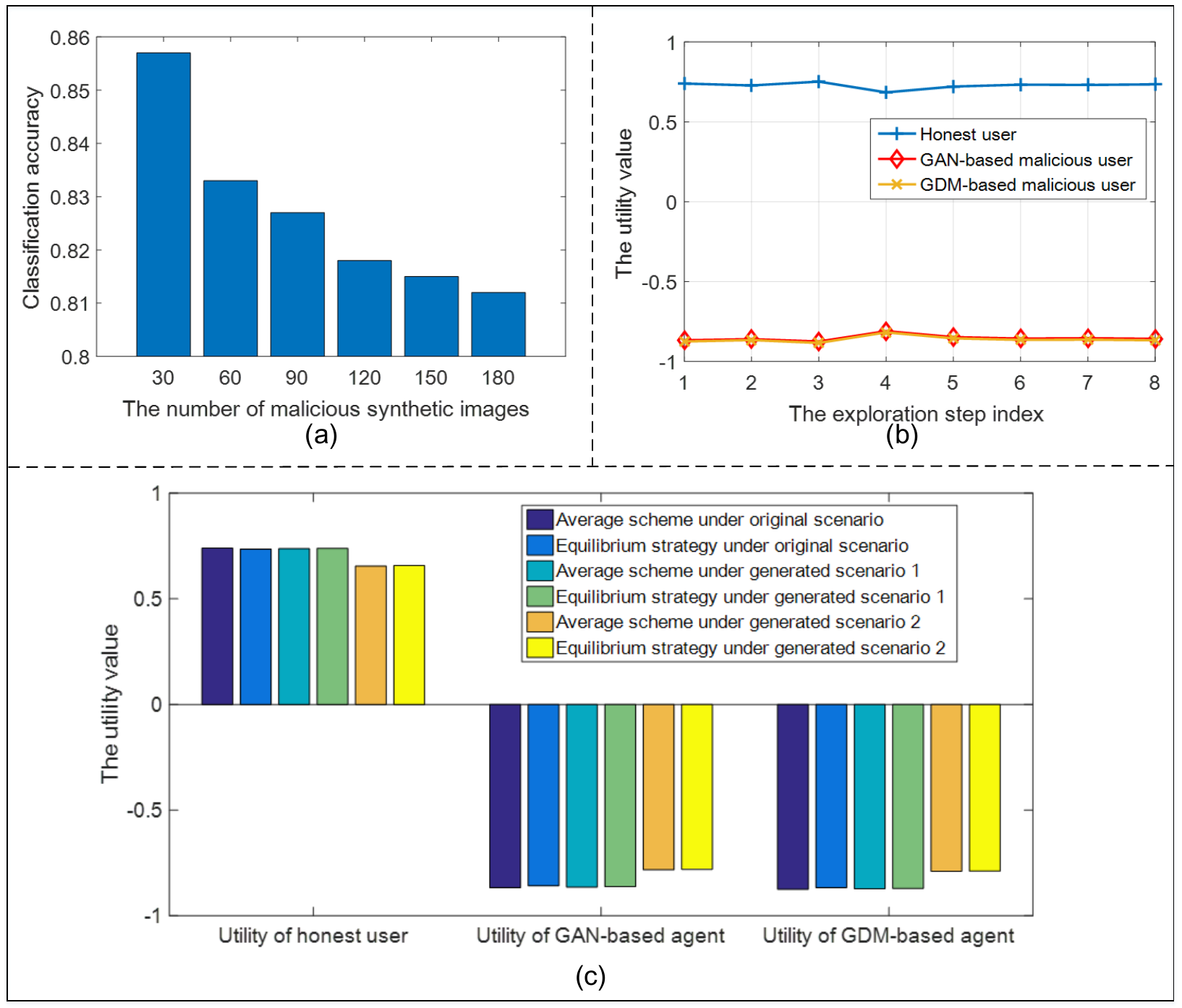}\\
  \caption{Performance evaluation for the proposed framework. (a) The ML model accuracy performance under different number of synthetic images generated by GAN-based agent and GDM-based agent. (b) The utility convergence process for honest user, GAN-based agent, and GDM-based agent. (c) Equilibrium strategy performance evaluation under newly generated game scenarios. \textcolor[rgb]{0.00,0.00,0.00}{Note that utility functions designed for honest and malicious users are (SVM model's accuracy -- collection cost) and (-SVM model's accuracy -- generation cost), respectively. The game strategy indicates the numbers of images collected or generated by honest and malicious users. The Nash equilibrium means that when other player's strategy remains unchanged, any player (i.e., honest user or malicious user) who unilaterally changes its strategy will not increase its profit. To be detailed, after finishing initial strategy exploration, we iteratively search out the Nash equilibrium by inputting the current exploration results to ChatGPT 4 to acquire the next step exploration direction. When the utilities of the players converge to be steady, the Nash equilibrium solution can be reached.}}
  \label{simulation}
\end{figure*}

\emph{Framework implementation}. In the input layer, we begin by inputting the optimization scenario and problem into ChatGPT 4, enabling the formulation of a theoretical game model within the conceptual/virtual space. Subsequently, we employ GAN and GDM as malicious users to generate synthetic/false images. In the processing layer, we initialize the number of images from honest users, GAN-based agents, and GDM-based agents to (150, 75, 75) images. Then, we train the SVM to determine model accuracy and pre-defined utilities at exploration step 1. Continuing, we calculate performances under strategies (180, 60, 60) and (120, 90, 90) at exploration steps 2 and 3, respectively. These results are inputted into ChatGPT 4 to obtain the next exploration strategy. For instance, steps 4 and 5 involve exploration strategies guided by ChatGPT 4, resulting in (135, 83, 82) and (142, 79, 79), respectively. Once utilities converge to steady points, they indicate equilibrium strategies. Moreover, we utilize GAN and GDM to generate another two sets of new images, simulating two new games\footnote{\textcolor{black}{Note we first utilize GAN and GDM to generate 200 synthetic images as initial dataset at one time, representing action sets of the original game. Then, we create two new games by changing their action sets, i.e., generating another two synthetic datasets for new game scenarios--generated scenarios 1 and 2.}}. Following these implementation details, the performance of equilibrium strategies can be evaluated under new generated game scenarios, with the strategy (150, 75, 75) serving as the baseline for the average scheme. Finally, in the output layer, the final results are obtained by reflecting the relationship between game strategies and optimization results.



\emph{Results analysis}. Fig. 2 assesses the performance of the proposed framework. Innitially, Fig. 2(a) illustrates that as the number of synthetic images increases, the ML accuracy performance decreases. This demonstrates the opponent modeling capacity of generative AI, where GAN and GDM can act as the attacker to provide synthetic/false data to degrade the ML model performance. Subsequently, Fig. 2(b) evaluates the feasibility of leveraging generative AI to create agents for exploring game results. As depicted, the utilities of honest user, GAN-based agent, and GDM-based agents all can convergence to stable equilibrium strategy as the exploration step increases. Moreover, Fig. 2(c) evaluates the performance of derived Nash equilibrium strategy under new generated game scenarios, where the utility of honest user, GAN-based agent, GDM-based agent remain optimal compared to the baseline average scheme.

\section{Future research directions}

\subsection{Game theory for generative AI}

Except generative AI can bring benefits to the whole life cycle of game theory, game theory can also provide significant insights to generative AI. For generative AI models employing adversarial training architecture, game theory serves as a guiding force, accelerating convergence speed, enhancing outcome stability, and refining architecture design. This synergy between generative AI and game theory is not merely complementary; it represents a mutually beneficial advancement, fostering symbiotic development.

\subsection{Real-time response requirement}

While generative AI offers effective solutions to game theoretical problems, its real-time application may encounter response delays due to extensive data training and complex architecture design. To address this challenge, lightweight model structures, parallel computing, and streaming and chunking techniques can be implemented in generative AI. These strategies  can mitigate response latency, enhancing its utility in latency-sensitive scenarios and interactive environments.

\subsection{Ethical and social implications}

Generative AI-enabled game theoretical approaches can deploy intelligent agents to supplant real-world individuals, societies, and organizations in strategic decision-making. Thus, ethical and social considerations must guide the design and regulation of agent behaviors and interaction rules. This approach ensures compliance with ethical norms and mitigates the risk of societal repercussions stemming from ethical oversights. Such conscientiousness fosters wider acceptance and adoption of generative AI-enabled game theoretical frameworks within human society.

\section{Conclusion}

In this paper, we delve into the integration of generative AI and game theory, shedding light on the necessity, advantages, and merits of incorporating generative AI into game theory frameworks. Specifically, we explore applications of generative AI-enhanced game theory across the entire spectrum of game theory, spanning model formulation, solution derivation, and strategy refinement. Furthermore, we introduce a generative AI-fortified game theoretical framework tailored to optimize machine learning model performance amidst false data injection attacks. Through a case study centered on mobile crowdsensing-based vehicle image collection, we showcase the feasibility and efficacy of our proposed framework through numerical results. \textcolor[rgb]{0.00,0.00,0.00}{Finally}, we outline several promising directions for future research.

\bibliography{ref}{}
\bibliographystyle{IEEEtran}

\end{document}